\documentclass[twocolumn]{aastex631}

\usepackage{amsmath}

\begin{document}

\title{Viscous Transonic Accretion Flows in Kerr Black Hole Geometry}

\email{abhrajitb12@gmail.com, sandipchakrabarti9@gmail.com}

\author{Abhrajit Bhattacharjee}
\affiliation{Indian Centre for Space Physics, 466, Barakhola, Netai Nagar, Kolkata 700099, India}

\author{Sandip K. Chakrabarti}
\affiliation{Indian Centre for Space Physics, 466, Barakhola, Netai Nagar, Kolkata 700099, India}



\begin{abstract}
We study viscous transonic accretion flows in vertical equilibrium in Kerr geometry. We employ the pseudo-Kerr formalism which accurately describes transonic flows around Kerr black holes and is applicable for modelling observational data. We study the effects of viscosity on the nature of sonic points and the parameter space that allows an accretion flow to possess multiple sonic points. We concentrate on the accretion solutions that can have centrifugal pressure supported shock waves and find that the shocks are weaker and are located farther from the black hole as the viscosity is enhanced. Moreover, if the viscosity is greater than a critical value, shocks do not form and the accretion flow can pass only through the inner sonic point close to the black hole and remains subsonic and Keplerian throughout the accretion disk. Since the resonance oscillation frequencies of the shock waves provide a measure of the observed Quasi Periodic Oscillation (QPO) frequencies, and since the location of shock waves depend on the spin of a black hole, it is clear that the QPO frequencies must depend on the spin of black hole as well. Our pseudo-Kerr approach makes it easier to compute spectra from an accretion flow with viscous dissipation and radiative cooling around a spinning black hole.
\end{abstract}



\section{Introduction} \label{sec:intro}
The accretion of matter onto black holes is a ubiquitous mechanism that is believed to be associated with a diverse array of high-energy astrophysical phenomena \citep{pringle1981accretion, frank2002accretion}. It is well known that the standard disk models of \citet{shakura1973black} and \citet{novikov1973astrophysics} could explain only the thermal component of the black hole accretion disk spectrum and unable to account for the non-thermal power-law component that is usually present in a general observed spectrum \citep{sunyaev1979hard}. These models assume that the entire flow is subsonic and that the disk is terminated at the marginally stable orbit. Moreover, while ignoring the inner boundary condition at the horizon, the advection and pressure-gradient terms were also not treated properly. It was later realised that for any generic equation of state, the flow must be supersonic close to the horizon and the matter velocity attains the velocity of light while on the horizon. Consequently, the flow must be sub-Keplerian at the inner regions of the accretion disk, leading to a deviation from the standard Keplerian disk \citep{chakrabarti1996accretion}. Thus, black hole accretion process is necessarily  transonic  \citep{liang1980transonic, chakrabarti1990theory}. \par 

It was suggested by \citet{sunyaev1980comptonization, sunyaev1985comptonization} that any model having a Comptonizing hot electron cloud in addition to the standard Keplerian disk can explain the power-law component of the emitted spectrum observed especially in the hard state. The two-component (Keplerian and sub-Keplerian) advective flow (TCAF) solution of \citet[hereafter CT95]{chakrabarti1995spectral} while successfully addressing the shortcomings of the standard disk, models the accretion flow as a Keplerian disk at the equatorial plane immersed inside a sub-Keplerian (low angular momentum) accreting halo which produces a centrifugal-barrier-dominated hot region close to the black hole. This hot inner disk region behaves like a Compton cloud of hot electrons that inverse comptonizes the soft photons produced by the Keplerian disk \citep{ghosh2009monte}, thus producing the power-law spectral slope in the soft states of black hole candidates. This region is also considered to be responsible for producing outflows and jets \citep{chakrabarti1999estimation, mondal2021spectral}. Furthermore, the numerical simulations of \citet{giri2013hydrodynamic} reveal that the TCAF configuration is not only achievable but also stable. The TCAF solution has also been applied, with certain modifications, to weakly magnetized neutron stars as well \citep{bhattacharjee2017monte, bhattacharjee2019timing}. \par

The existence of shocks and their astrophysical implications in the context of black hole accretion flows has been thoroughly investigated following the revelation that a rotating flow features more than one sonic point \citep{liang1980transonic}. It is well-known that a transonic flow may undergo shock transitions when the Rankine-Hugoniot shock conditions are satisfied. \citet{chakrabarti1989standing} carried out a detailed analysis of the shock solutions of inviscid, transonic flows and classified the parameter space in terms of whether the Rankine-Hugoniot shock conditions are satisfied or not, as a subclass of the solutions that allows multiple sonic points, both in accretion and in winds. The analysis was later extended to viscous transonic flows \citep[hereafter C96b]{chakrabarti1996grand}. Subsequently, time-dependent numerical simulations indicated that these shocks are stable as well \citep{chakrabarti1993smoothed, molteni1994simulation, molteni1999azimuthal}. All these theoretical investigations employed the pseudo-Newtonian potential of \citet[hereafter PW80]{paczynsky1980thick} and was later generalized to full general relativity \citep{chakrabarti1996global}. The PW80 potential has been used in numerous works to investigate the physical properties of accretion flows around non-rotating black holes \citep{chakrabarti1996accretion, chattopadhyay2002radiatively, proga2003accretion, beckwith2011turbulence, singh2012nature, mondal2013spectral}. General relativistic numerical simulations also demonstrate that stable shocks can indeed form in accretion flows around black holes \citep{sukova2017shocks, kim2017general, kim2019general}. It was shown later that the resonance oscillation of the shocks are the driving force behind the so-called quasi-periodic oscillations (QPOs) observed from black hole candidates \citep[see][and references therein]{chakrabarti2015resonance}. Numerical simulations of magnetized accretion flows have also been performed to understand the effect of magnetic field on the shock \citep{deb2017dynamics, okuda2019possible, garain2020effects}.\par

The Kerr solution describes in general relativity, the spacetime of astrophysically relevant black holes that are solely characterized by just two parameters - gravitational mass $M$ and angular momentum $J$ (usually defined by the spin parameter $a\equiv J/M$). Black hole spin introduces qualitatively new features into gravitational dynamics that have significant effects in models of accretion disks around black holes. In an earlier work \citep{bhattacharjee2022transonic}, we studied inviscid, transonic flows around rotating black holes using a pseudo-Kerr effective potential that mimicked the behavior of Kerr geometry and discussed the formation of shocks both in accretion and in winds. Although the conditions around black holes are extreme, the pseudo-Kerr formalism affords an excellent approximation as evidenced by comparison with results obtained using full general relativity \citep{chakrabarti1996global}. Having satisfied ourselves with the accuracy of the formalism, we extend our analysis in this paper to study the effects of viscosity on accretion flows in Kerr geometry. We adopt the viscosity prescription of \citet{chakrabarti1995viscosity} where the viscous stress $W_{r\phi}$ is equated to $-\alpha_\Pi(W + \Sigma v^2)$, where $\alpha_\Pi$ is the viscosity parameter, $W$ and $\Sigma$ are the vertically integrated pressure and matter densities, respectively. This prescription is particularly important when the radial velocity of the flow is significant and this ensures that the angular momentum and viscous stress are continuous across shock waves. \par

The plan of this paper is as follows: In the next section, we present the basic equations governing viscous transonic accretion flows in the equatorial plane of a Kerr black hole. In Section 3,  we carry out the sonic point analysis. In Section 4, we study shock solutions in viscous accretion flows around Kerr black holes and in Section 5, we discuss our results. Finally, in Section 6, we present the concluding remarks. \par

\section{Basic Flow Equations} \label{sec:bfe}
We consider a viscous, stationary and axisymmetric accretion flow around a rotating black hole described by the Kerr metric. We choose the geometric units $G=M=c=1$ ($G$ is the gravitational constant, $M$ is the mass of the black hole and $c$ is the speed of light) such that the units of velocity, distance and time are $c$, $GM/c^2$ and $GM/c^3$ respectively. \par

We choose cylindrical coordinates ($t,r,\phi,z$) and assume the flow to be vertically averaged. The local half-thickness of the flow, $H(r)$, obtained by equating the pressure gradient force in the vertical direction with the component of the gravitational force along that direction, is assumed to be much smaller than the cylindrical radial coordinate, i.e., $H(r)<<r$. We assume the vacuum metric in and near the equatorial plane of a Kerr black hole to be of the form \citep{novikov1973astrophysics}
\begin{equation}
\begin{split}
    ds^2 & = g_{\mu\nu}dx^\mu dx^\nu \\
         & = -\frac{r^2\Delta}{A}dt^2 + \frac{A}{r^2}(d\phi-\omega dt)^2 + \frac{r^2}{\Delta}dr^2 + dz^2,
\end{split}
\end{equation}
where $A=r^4+r^2a^2+2ra^2$, $\Delta=r^2-2r+a^2$, $\omega=2ar/A$ and $a$ is the spin parameter of the black hole. Here, $g_{\mu\nu}$ is the metric coefficient and the four-velocity components $u_\mu$ satisfies the normalization condition $u_\mu u^\nu=-1$ which is a conserved quantity. The event horizon of the black hole is located at the outer root of $\Delta=0$, i.e., $r_h=1+\sqrt{1-a^2}$. The self-gravity of the flow is ignored and the central plane of the accretion disk is assumed to be aligned with the equatorial plane ($z=0$) of the black hole. \par

The matter distribution is assumed to be described by the perfect fluid stress-energy tensor
\begin{equation}
    T_{\mu\nu} = h\rho u_\mu u_\nu + pg_{\mu\nu},
\end{equation}
which satisfies the equations of motion 
\begin{equation}
    \nabla_\mu T^{\mu\nu} = 0.
\end{equation}
Here, $h=1+\epsilon+P/\rho$ is the specific enthalpy, $P$ is the isotropic pressure, $\rho$ is the rest-mass density and $\epsilon$ being the specific internal energy defined in the local rest frame of the fluid. \par

The radial velocity $v$ in the corotating frame, a frame that rotates with the same angular velocity as the flow, is given by
\begin{equation}
    v = \left(1 + \frac{1}{g_{rr}u^r u^r}\right)^{-1/2}.
\end{equation}
Then, one has $|v|\leq 1$ everywhere in the flow and $|v|=1$ at the event horizon, independent of the mass and spin of the black hole. Since the local sound speed is always much less than unity even for the extreme equation of states, the black hole accretion process must be supersonic (and therefore sub-Keplerian) at the event horizon. Thus, any flow must deviate from Keplerian nature close to the event horizon of the black hole and the flow must be transonic in nature \citep{chakrabarti1990theory}. \par

In what follows, we concentrate on the stationary solutions of the underlying hydrodynamics equations. We shall use the viscosity prescription of \citet{chakrabarti1995viscosity} which is appropriate for studying flows with significant radial motion:
\begin{equation}
    W_{r\phi} = -\alpha_\Pi \Pi = -\alpha_\Pi(W + \Sigma v^2).
\end{equation}
Here, $W$ and $\Sigma$ are the vertically integrated pressure and matter densities \citep{matsumoto1984viscous}
\begin{equation}
    \Sigma = \int^{h_0(r)}_{-h_0(r)}\rho(r,z)dz = 2\rho I_n h_0(r)
\end{equation}
and 
\begin{equation}
    W = \int^{h_0(r)}_{-h_0(r)}p(r,z)dz = 2p I_{n+1} h_0(r),
\end{equation}
where $I_n=(2^n n!)^2/(2n+1)!$, $n=1/(\gamma-1)$ is the polytropic index and $h_0(r)$ is the local half-thickness of the flow. Since the total pressure $\Pi = W + \Sigma v^2$ is continuous across a shock, this $\Pi-$stress prescription ensures that the viscous stress $W_{r\phi}$ as well as the angular momentum of the flow is continuous across the shock. \par

\begin{figure}
    \plotone{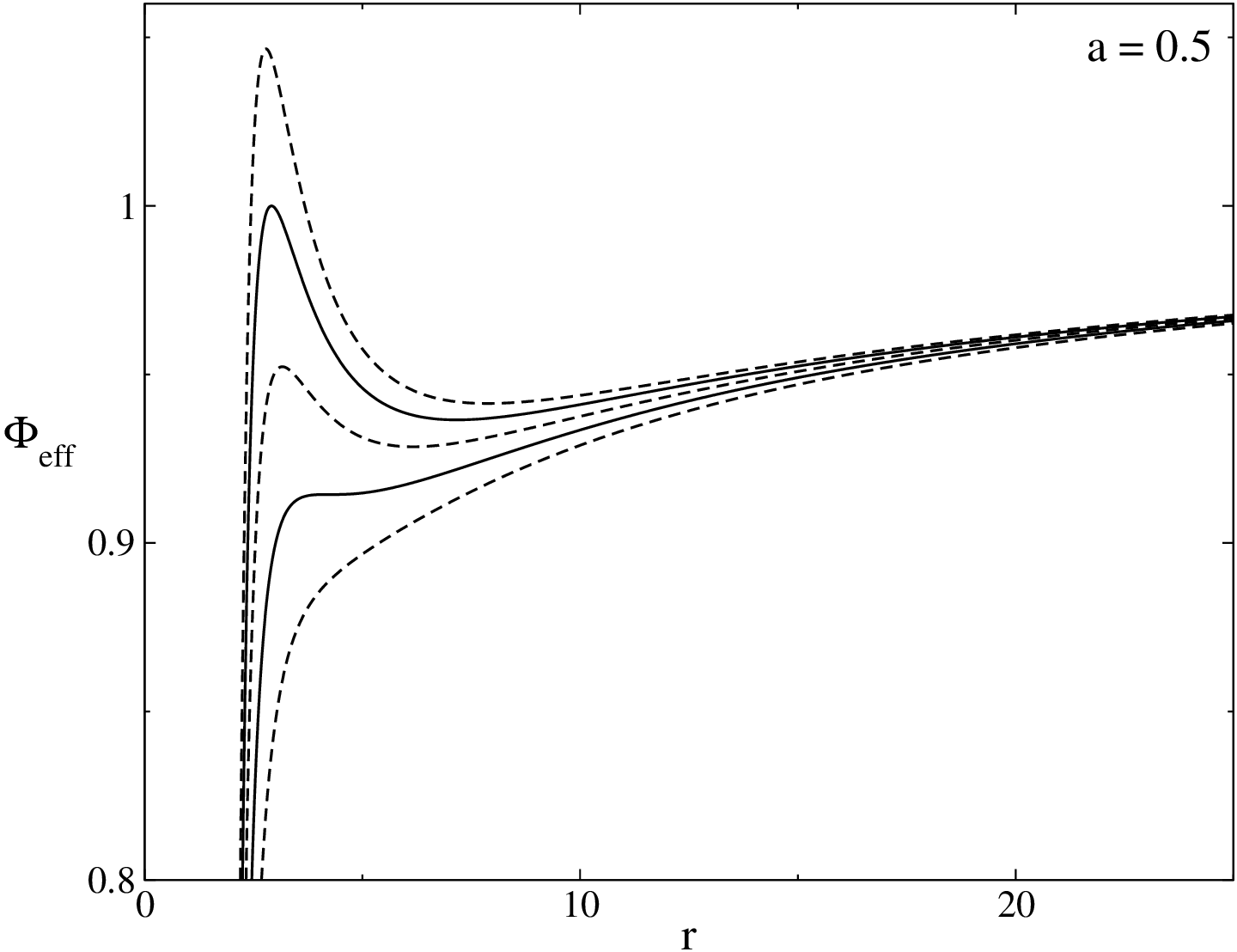}
    \caption{The effective potential $\Phi_\mathrm{eff}$ for $a=0.5$ and $l=3.0, 3.1629, 3.3, 3.4142, 3.5$ (from the lowermost curve upwards). The lower solid curve represent the marginally stable angular momentum and the upper solid curve represent the marginally bound angular momentum.}
\end{figure}

The radial momentum equation is obtained from the projection of Equation (3) onto the space orthonormal to the four-velocity using the projection tensor $h_{\mu\nu} = g_{\mu\nu} + u_\mu u_\nu$:
\begin{equation}
    h\rho u^\mu\nabla_\mu u^r + (g^{\mu r} + u^\mu u^r)\partial_\mu p = 0.
\end{equation}
This equation can be written in the corotating frame, using Equation (4), as follows:
\begin{equation}
    v\frac{dv}{dr} + \frac{1}{\rho}\frac{dp}{dr} + \frac{d\Phi_\mathrm{eff}}{dr} = 0,
\end{equation}
where
\begin{equation}
    \Phi_\mathrm{eff} = 1 + \frac{1}{2}\ln\left(\frac{r\Delta}{r^3+(a^2-l^2)r+2(a-l)^2}\right)
\end{equation}
is the pseudo-Kerr effective potential \citep{bhattacharjee2022transonic} calculated at the equatorial plane of a Kerr black hole. The specific angular momentum of the flow is defined as $l=-u_\phi/u_t$, where $u_\phi$ and $u_t$ are the azimuthal component of the four-velocity and the specific binding energy, respectively. Here, we assume that for all practical situations, the Lorentz factor $\gamma_v=1$ and specific enthalpy $h\sim 1$ all throughout. \par  

Fig. 1 shows the effective potential $\Phi_\mathrm{eff}$ as a function of the radial coordinate for different values of the specific angular momentum and the spin parameter is chosen to be $a=0.5$. The solid curves represents the marginally stable ($l_{ms}=3.1629$; lower curve) and the marginally bound ($l_{mb}=3.4142$; upper curve) cases and the dashed curves represents $l=3.0, 3.3$ and $3.5$ (from bottom to top). Now, Equation (9) can be expressed in terms of the adiabatic sound speed $a_s (=\sqrt{\gamma P/\rho})$ as 
\begin{equation}
    v\frac{dv}{dr} + \frac{2a_s}{\gamma}\frac{da_s}{dr} + \frac{a_s^2}{\gamma\rho}\frac{d\rho}{dr} = -\frac{d\Phi_\mathrm{eff}}{dr}.
\end{equation}
The continuity equation can be expressed in the form
\begin{equation}
    \frac{1}{v}\frac{dv}{dr} + \frac{1}{a_s}\frac{da_s}{dr} + \frac{1}{\rho}\frac{d\rho}{dr} = -G,
\end{equation}
where 
$$G=h'(r)/h(r), \quad h(r)=(r/a_s)h_0(r)$$
and
\begin{equation}
h_0(r)=a_sr^{1/2}(\Phi')^{-1/2}
\end{equation}
is the local half-thickness of the flow obtained by equating the pressure gradient force and the force due to the gravitational potential $\Phi=\Phi_\mathrm{eff}(a,r,l=0)$ along the vertical direction, thus assuming vertical hydrostatic equilibrium. Hereafter, we shall use prime to represent derivative with respect to the radial coordinate $r$ in flat geometry. \par

The equation for the conservation of angular momentum is
\begin{equation}
    v\frac{dl}{dr} + \frac{1}{\Sigma r}\frac{d}{dr}(r^2 W_{r\phi}) = 0,
\end{equation}
where the vertically integrated viscous stress $W_{r\phi}$ keeps the angular momentum continuous across any shock transition in the flow. This equation can be integrated to yield
\begin{equation}
    l-l_{in}=\frac{\alpha_\Pi r}{v\gamma}(ga_s^2 + \gamma v^2).
\end{equation}
Here, $g=I_{n+1}/I_n$ and $l_{in}$ is the angular momentum at the inner edge of the accretion disk. It is evident that for an inviscid flow ($\alpha_\Pi=0$), one recovers $l=l_{in}(=\textit{constant})$ which is a conserved quantity. \par

After some algebra, the differential form of Equation (14) may be written as 
\begin{equation}
    \frac{dl}{dr} = \alpha_\Pi\left[v\left(1+\frac{ga_s^2}{\gamma v^2}\right) +\left(1-\frac{ga_s^2}{\gamma v^2}\right)\frac{dv}{dr}+\frac{2ga_s r}{\gamma v}\frac{da_s}{dr}\right].
\end{equation}

The entropy generation equation is (C96b)
\begin{equation}
    \Sigma vT\frac{ds}{dr} = \frac{h_0v}{\Gamma_3-1}\left[\frac{dp}{dr}-\Gamma_1\frac{p}{\rho}\frac{d\rho}{dr}\right] = Q^+ - Q^-,
\end{equation}
where
$$\Gamma_1 = \frac{\beta+(4-3\beta)^2(\gamma-1)}{\beta+12(\gamma-1)(1-\beta)},\quad \Gamma_3=1+\frac{\Gamma_1-\beta}{4-3\beta}$$
and 
$$\beta = \frac{\rho kT/\mu m_p}{\bar{a}T^4/3+\rho kT/\mu m_p}$$
is the ratio of gas pressure to total pressure ($\bar{a}$, $k$, $m_p$ and $\mu$ are the Stefan constant, the Boltzmann constant, mass of proton and mean molecular weight, respectively), $s$ is the entropy density of the flow, $T$ is the total temperature, $Q^+$ and $Q^-$ are the heat-generation  rate and heat-loss rate, respectively. In the present analysis, we shall use $\Gamma_1=\gamma=\Gamma_3$ and ignore cooling effects explicitly (i.e., $Q^-=0$). We shall assume an equation of state valid for an ideal gas so that Equation (17) can be expressed as 
\begin{equation}
    \frac{va_s^2}{\gamma}\left[\frac{2n}{a_s}\frac{da_s}{dr}-\frac{1}{\rho}\frac{d\rho}{dr}\right] = -\frac{Q^+}{h_0\rho} = -H.
\end{equation}

The heating rate $Q^+$ is calculated using the MIxed Shear Stress (MISStress) prescription of C96b in which two forms of the viscous shear stress, $W_{r\phi}^{(1)} = -\alpha_\Pi \Pi$ and $W_{r\phi}^{(2)} = \eta r\frac{d\Omega}{dr}$, are used in the following way:
\begin{equation}
    Q^+ = \frac{W_{r\phi}^2}{\eta} = \frac{W_{r\phi}^{(1)}W_{r\phi}^{(2)}}{\eta}.
\end{equation}

This equation can be used to write the heating rate $H$ as
\begin{equation}
    H = A_n(ga_s^2 + \gamma v^2)r\frac{d\Omega}{dr},
\end{equation}
where $A_n=-\alpha_\Pi I_n/\gamma$, $g=I_{n+1}/I_n$ and $\Omega(r)=l(r)/r^2$ is the angular velocity of the accreting matter. \par

Using Equation (16) in Equation (20), we obtain
\begin{multline}
    H = A_n\Theta\left(\frac{\alpha_\Pi\Theta}{\gamma vr}-\frac{2l}{r^2}\right) -\frac{\alpha_\Pi A_n}{\gamma v^2}(g^2a_s^4-\gamma^2v^4)\frac{dv}{dr} \\
    + 2\alpha_\Pi A_n\Theta\frac{ga_s}{\gamma v}\frac{da_s}{dr},
\end{multline}    
where $\Theta = ga_s^2 + \gamma v^2$ is defined for simplicity. Hence, Equation (18) becomes
\begin{multline}
    A_n\alpha_\Pi\frac{\gamma^2v^4-g^2a_s^4}{v^3a_s^2}\frac{dv}{dr} + \left(\frac{2n}{a_s}+\frac{2A_n\alpha_\Pi g\Theta}{v^2a_s}\right)\frac{da_s}{dr} \\
    = \frac{1}{\rho}\frac{d\rho}{dr} -A_n\Theta\left(\frac{\alpha_\Pi\Theta}{v^2a_s^2r}-\frac{2\gamma l}{va_s^2r^2}\right).
\end{multline}

Along with the equations given above, one requires to solve the Rankine-Hugoniot shock conditions if the flow passes through a shock transition.

\section{Analysis of the Sonic Points} \label{sec:asp}

\begin{figure*}
    \epsscale{0.8}
    \plotone{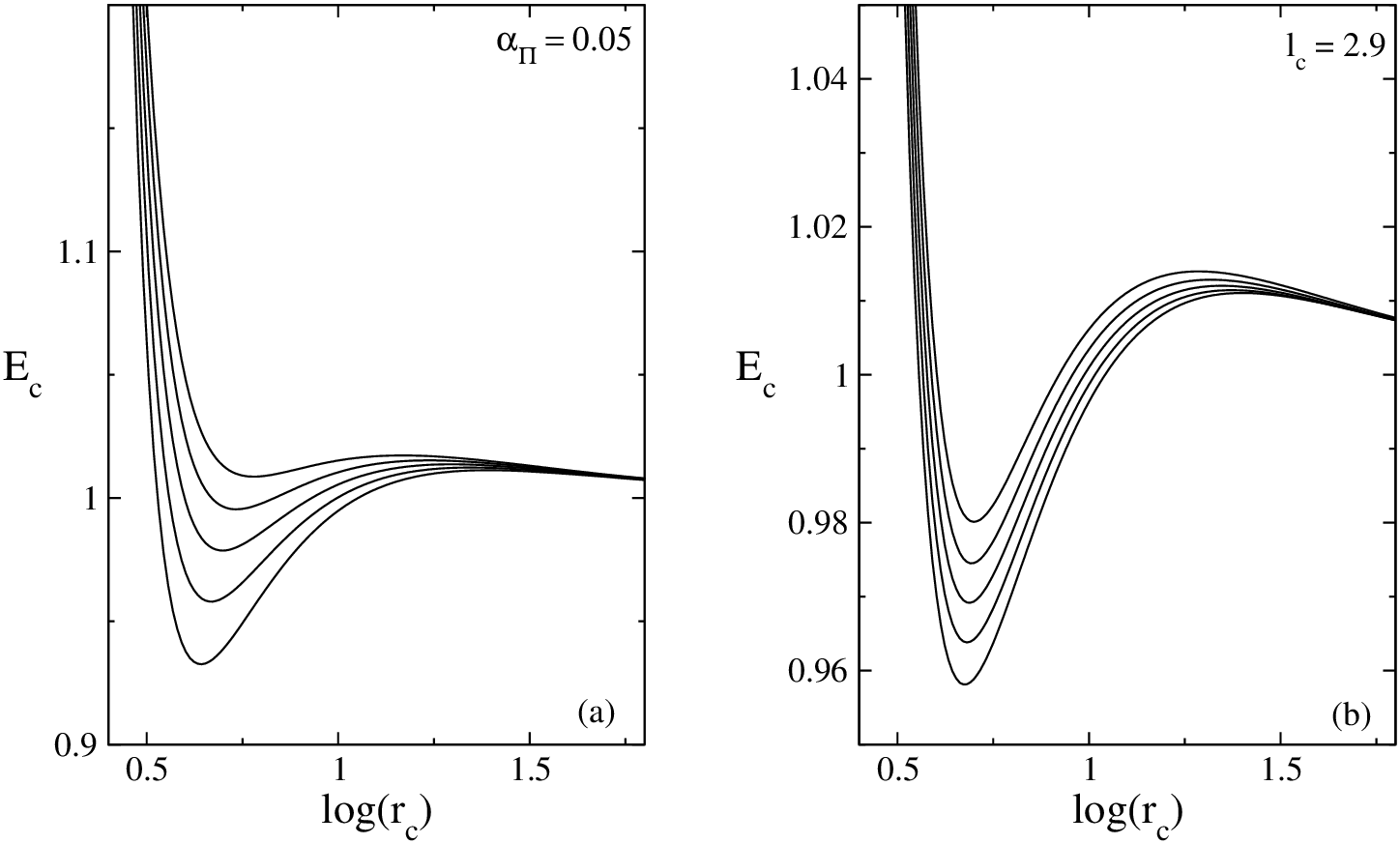}
    \caption{Variation of the specific energy of the accretion flow as a function of the location of the sonic points for (a) $\alpha_\Pi=0.05$ and $l_c=2.7, 2.8, 2.9, 3.0, 3.1$ (from the uppermost curve downwards) and (b) $l_c=2.9$ and $\alpha_\Pi=0, 0.2, 0.4, 0.6, 0.8$ (from the uppermost curve downwards). The spin parameter is $a=0.5$ for both the plots.}
\end{figure*}

We begin by expressing the governing equations of the flow, namely Equations (11), (12), (15) and (22), in matrix form as follows:
\begin{multline}
\begin{pmatrix}
    \dfrac{A_n\alpha_\Pi\Theta(\Theta-2ga_s^2)}{a_s^2v^3}  & \left[\dfrac{2n}{a_s}+\dfrac{2A_n\alpha_\Pi g\Theta}{a_sv^2}\right] & -1 \\
    \dfrac{1}{v} & \dfrac{1}{a_s} & 1 \\
    v & \dfrac{2a_s}{\gamma} & \dfrac{a_s^2}{\gamma}
\end{pmatrix} \\
\begin{pmatrix}
    \dfrac{dv}{dr} \\
    \dfrac{da_s}{dr} \\
    \dfrac{d\log\rho}{dr}
\end{pmatrix}
=
\begin{pmatrix}
    -A_n\Theta\left[\dfrac{\alpha_\Pi\Theta}{v^2a_s^2r}-\dfrac{2(n+1)l}{na_s^2vr^2}\right] \\
    -G \\
    -\Phi'_\mathrm{eff}
\end{pmatrix}
\end{multline}

After some algebra, using Cramer's rule, we obtain
\begin{equation}
    \frac{dv}{dr}=\frac{N}{D},
\end{equation}
where
\begin{equation}
\begin{split}
    N & = \left[(2n+1)v^2+2A_n\alpha_\Pi g\Theta\right]\Phi'_\mathrm{eff} + \frac{2A_nlv}{r^2}\Theta \\ 
    & - \frac{A_nn\alpha_\Pi}{(n+1)r}\Theta^2 - \frac{2A_nn\alpha_\Pi ga_s^2}{n+1}G\Theta - 2na_s^2v^2G
\end{split}
\end{equation}
and
\begin{equation}
\begin{split}
    D & = 2na_s^2v - (2n+1)v^3 \\ 
    & - A_n\alpha_\Pi v\Theta\left[(2g-1)-\frac{gna_s^2}{(n+1)v^2}\right].
\end{split}
\end{equation}

Since the radial velocity of the matter is negligible at the outer edge of an accretion disk at large radii, the flow is subsonic ($v<a_s$). On the contrary, the flow is supersonic ($v>a_s$) near the horizon, in particular the matter enters the event horizon with the speed of light. This means that the denominator ($D$) of Equation (24) must vanish at an intermediate location and to have a smooth solution across that location the numerator ($N$) must also vanish simultaneously. Such a location defined by $N=D=0$ is the critical point ($r_c$) or sonic point of the flow where, by definition, $dv/dr$ is well-defined and regular. This means that the accretion flow is transonic in nature and the flow must have at least one sonic point. However, the formation of a standing shock demands that the angular momentum of the flow be significant so that at least two saddle-type sonic points are formed and the Rankine-Hugoniot shock conditions are satisfied in between them. \par

\begin{figure*}
    \epsscale{0.8}
    \plotone{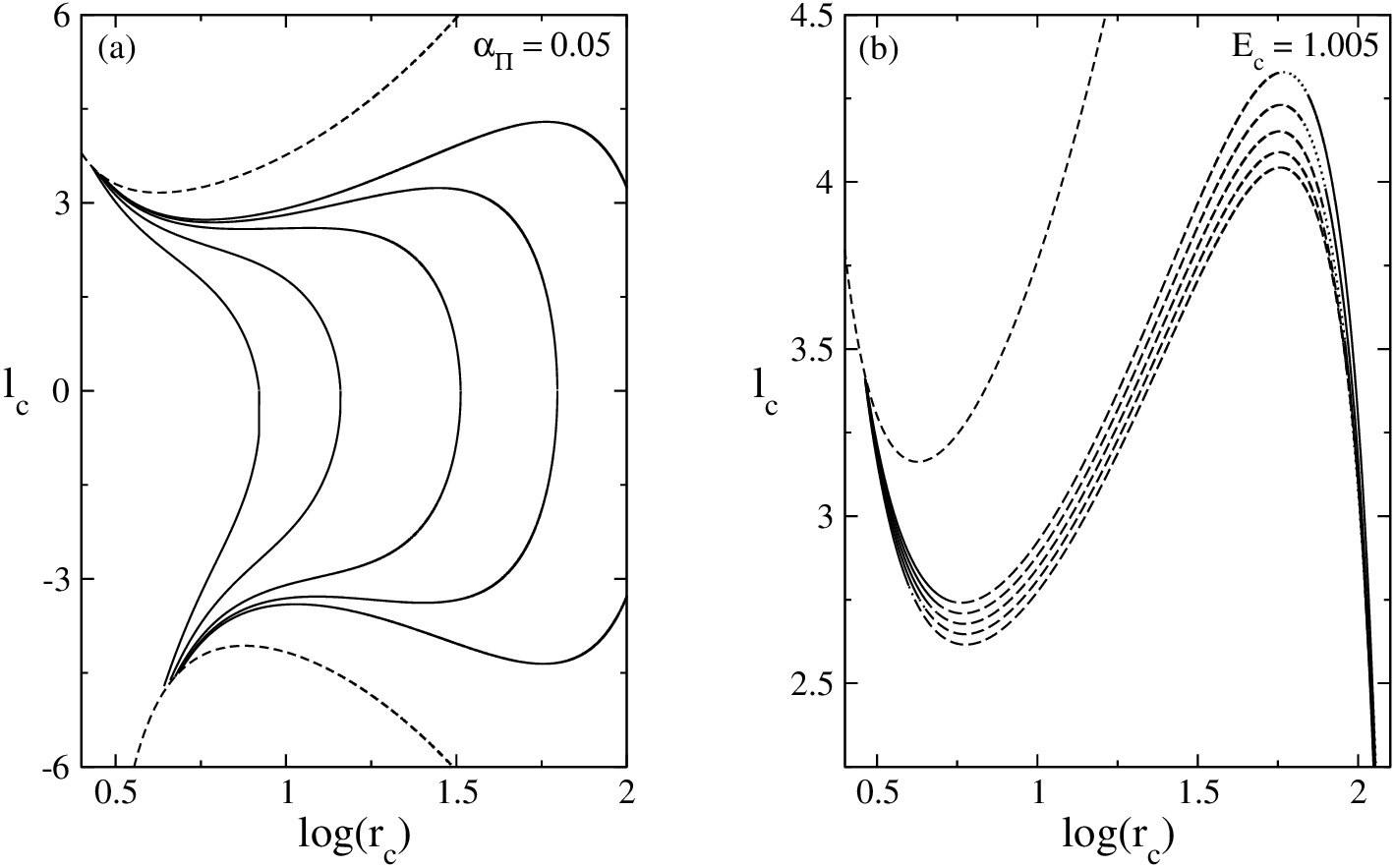}
    \caption{(a) Variation of the specific angular momentum ($l_c$) of the flow as a function of the location of the sonic points for $a=0.5$ and $\alpha_\Pi=0.05$ and different values of the specific energy at the sonic point, $E_c=1.003, 1.004, 1.005, 1.006, 1.007$ (from right to left). Prograde and retrograde flows correspond to $l_c>0$ and $l_c<0$, respectively. (b) Variation of $l_c$ as a function of the location of the sonic point is shown for $a=0.5$ and $E_c=1.005$ for different values of the viscosity parameter, $\alpha_\Pi=0, 0.15, 0.30, 0.45, 0.60$ (from top to bottom). Only the prograde flows are considered. The saddle-type, nodal-type and spiral-type sonic points are represented by the solid, dotted and long-dashed curves, respectively. In both the plots, the Keplerian distribution is shown by the short-dashed curves.}
\end{figure*}

Equating $D$ to zero, we obtain the expression for the Mach number $M=v/a_s$ at the sonic point:
\begin{equation}
    M_c = \frac{v_c}{a_{sc}} = \sqrt{\frac{-m_b-\sqrt{m_b^2-4m_am_c}}{2m_a}},
\end{equation}
where
\begin{equation*}
\begin{split}
    m_a & = -(2n+1)n - A_n\alpha_\Pi (n+1)(2g-1), \\
    m_b & = 2n^2 - 2A_n\alpha_\Pi ng(g-1), \\
    m_c & = \frac{A_n\alpha_\Pi n^2g^2}{n+1}.
\end{split}
\end{equation*}

In the weak-viscosity limit ($\alpha_\Pi\rightarrow 0$), the above expression reduces to
\begin{equation}
    M_c \approx \sqrt{\frac{2n}{2n+1}}
\end{equation}
as already obtained for inviscid flows using the pseudo-Kerr formalism \citep{bhattacharjee2022transonic}. This is of the same form even in full GR for flows in vertical equilibrium \citep{chakrabarti1996global}. It was shown in the context of thin inviscid flows that a small acoustic perturbation propagate with velocities $a_s[2n/(2n+1)]^{1/2}\pm v$ and so the true definition of the Mach number is $M=[(2n+1)/2n]^{1/2}(v/a_s)$ when the flow is in vertical equilibrium \citep{chakrabarti1989standing}. However, this redefinition of the Mach number has no impact on the following results and so we continue to use $M=v/a_s$ as the Mach number. \par 

Equating $N$ to zero, we obtain a transcendental equation for the sound speed at the sonic point of the form:
\begin{multline}
    \left[(2n+1)M_c^2 + 2A_n\alpha_\Pi g\tilde\Theta\right]\Phi'_\mathrm{eff} - 2nM_c^2Ga_{sc}^2 \\
    + \frac{2A_nlMa_{sc}\tilde\Theta}{r^2} - \frac{A_nn\alpha_\Pi a_{sc}^2\tilde\Theta}{n+1}\left[2Gg+\frac{\tilde\Theta}{r}\right] = 0,
\end{multline}
where we define $\tilde\Theta=\Theta/a_{sc}^2$. Note that the angular momentum $l$ is a function of the sound speed and the flow velocity (see Equation 15). We solve this equation numerically to obtain the sound speed at the sonic point. The angular momentum at the sonic point can be calculated using Equation (15) and the specific energy at the sonic point is
\begin{equation}
    E = \frac{v^2}{2}+\frac{a_s^2}{\gamma-1}+\Phi_\mathrm{eff}.
\end{equation}

The variation of the specific energy of the flow at the sonic point is shown in Fig. 2 for $a=0.5$. The curves in Fig. 2a are drawn for different values of the specific angular momentum at the sonic point, $l_c=2.7, 2.8, 2.9, 3.0, 3.1$ (from the uppermost curve to lowermost curve), for the viscosity parameter $\alpha_\Pi=0.05$. At large distances, the curves merge asymptotically to unity, i.e., the rest mass energy of the accretion flow. Notice that, for very low value of $l_c$, the flow may have only one sonic point close to the black hole. However, with increase in $l_c$, multiple sonic points may exist depending on the specific energy of the flow. The extreme locations of the inner and the outer sonic points can be determined from the minima and the maxima of the curves, respectively. In Fig. 2b, we show the variation of the specific energy for different values of the viscosity parameter, $\alpha_\Pi=0, 0.2, 0.4, 0.6, 0.8$ (from the uppermost curve to the lowermost curve), for fixed specific angular momentum at the sonic point ($l_c=2.9$). \par 

\begin{figure*}
    \epsscale{0.85}
    \plotone{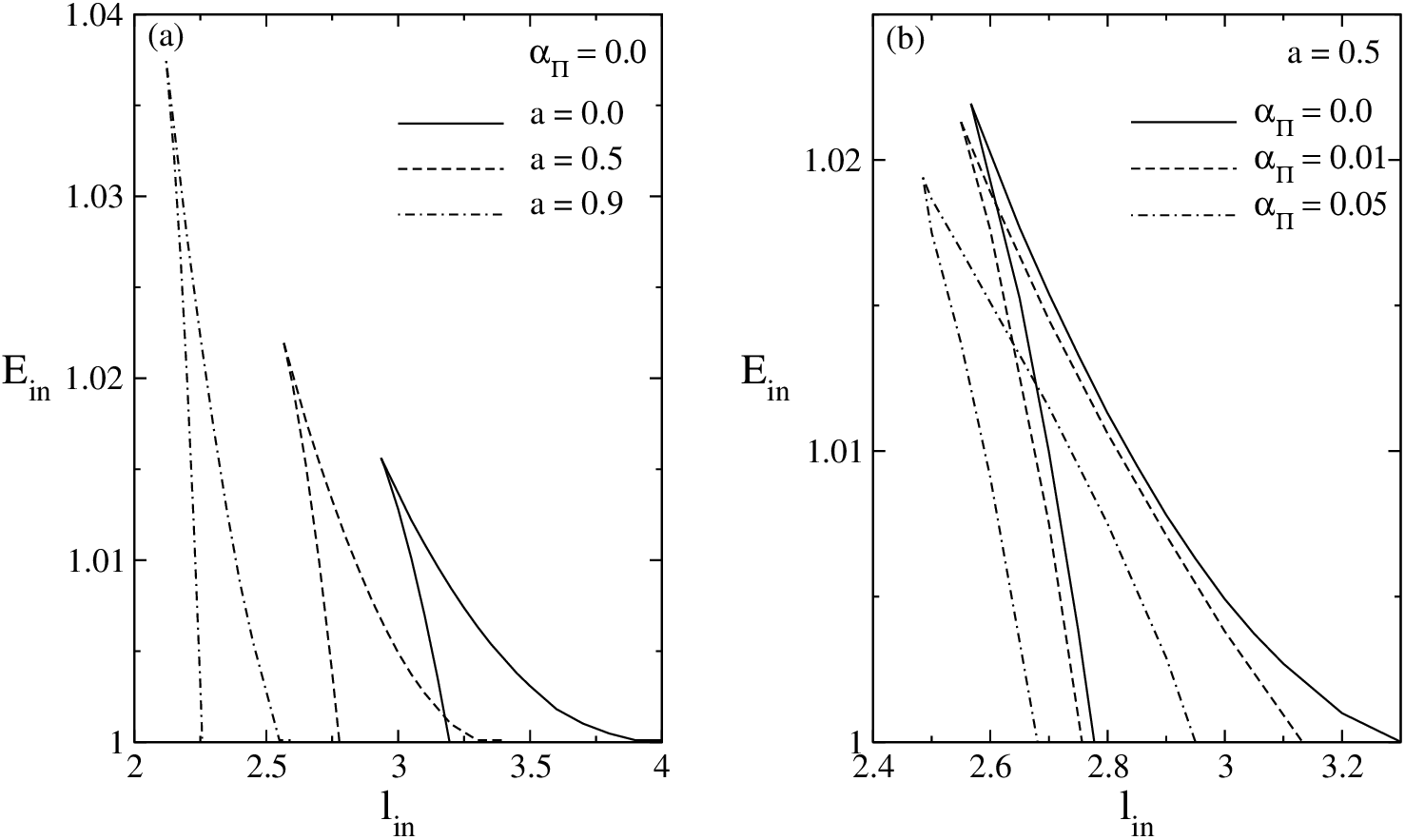}
    \caption{The parameter space spanned by the specific energy at the inner sonic point and the specific angular momentum of the accretion flow at the inner edge of the disk for (a) $\alpha_\Pi=0$ and different values of the spin parameter $a=0, 0.5, 0.9$ and (b) $a=0.5$ and different values of the viscosity parameter $\alpha_\Pi=0, 0.01, 0.05$. The bounded regions contain the parameters of the allowed solutions that may pass through the inner sonic point. In the inviscid case, both the specific energy and the specific angular momentum are conserved throughout the accretion disk.}
\end{figure*}

Depending on the viscosity parameter $\alpha_\Pi$ and the initial parameters, namely the specific energy $E_{in}$ and the specific angular momentum $l_{in}$ at the inner edge of the accretion disk, a flow may possess multiple sonic points. At the sonic points, the radial velocity gradient is of the form $dv/dr=0/0$. So we apply the l'Hospital rule to calculate $(dv/dr)_c$ at the sonic points that can be expressed as a quadratic equation:

\begin{equation}
    \mathcal{A}\left(\frac{dv}{dr}\right)_c^2 + \mathcal{B}\left(\frac{dv}{dr}\right)_c + \mathcal{C} = 0, \\
\end{equation}

where

\begin{align*}
\begin{split}
\mathcal{A} ={}& 6na_s^2 + (10n+7)v^2 + \frac{A_nn\alpha_\Pi}{n+1}[J_0J_4+2J_1J_2],
\end{split}\\
\begin{split}
\mathcal{B} ={}& 12na_s^2Gv - 2(4n+3)v\Phi_\mathrm{eff}' - 4(n+1)Gv^3 \\
& - \frac{6A_nl\Theta}{r^2} + \frac{4A_n(n+1)lgv^2}{nr^2} + \frac{2A_nn\alpha_\Pi\Theta^2}{(n+1)vr} \\ 
& + \frac{8A_n\alpha_\Pi gJ_1}{n+1}\left[na_s^2G-(n+1)\Phi_\mathrm{eff}'\right] \\
& + \frac{4A_n\alpha_\Pi Gg\Theta}{(n+1)v}\left[na_s^2-(n+1)v^2\right],
\end{split}\\
\begin{split}
\mathcal{C} ={}& -[(2n+1)v^2+2A_n\alpha_\Pi g\Theta]\Phi_\mathrm{eff}'' \\
& + 2na_s^2v^2(2G^2+G') - 4(n+1)Gv^2\Phi_\mathrm{eff}'  \\
& + \frac{4A_nlv\Theta}{r^3} - \frac{3A_nn\alpha_\Pi\Theta^2}{(n+1)r^2} \\
& - \frac{4A_ng[lv-\alpha_\Pi r^2(Gv^2-g\Phi_\mathrm{eff}')]}{nr^2[na_s^2G-(n+1)\Phi_\mathrm{eff}']^{-1}} \\
& + \frac{2A_nna_s^2\alpha_\Pi g}{n+1}(4G^2ga_s^2+G'\Theta).
\end{split}
\end{align*}

Here, we define, $J_k = (\Theta/v) - k(1+1/n)gv$. From the above quadratic equation, we can calculate $(dv/dr)_c$ at the sonic points as
\begin{equation}
    \left(\frac{dv}{dr}\right)_c = \frac{-\mathcal{B}\pm\sqrt{\mathcal{B}^2-4\mathcal{A}\mathcal{C}}}{2\mathcal{A}}.
\end{equation}

The nature of the sonic points depend on the radial velocity gradient of the flow that assumes two values: one of them is valid for accretion whereas the other is valid for wind. The sonic point is saddle-type if both the derivatives are real and of opposite sign, whereas it is nodal-type when the derivatives are real and of the same sign and it is spiral-type if the derivatives are imaginary. In order to form a standing shock in an accretion flow, the flow must possess more than one saddle-type sonic point. \par

In Fig. 3a, we show the variation of the specific angular momentum of the flow as a function of the location of the sonic points for different values of the specific energy at the sonic point, $E_c=1.003, 1.004, 1.005, 1.006, 1.007$ (from right to left), for $a=0.5$ and the viscosity parameter $\alpha_\Pi=0.05$. The cases of both prograde ($l_c>0$) and retrograde ($l_c<0$) accretion flows are shown. The dashed curve represent the Keplerian angular momentum distribution that depends only on the spin parameter of the black hole and is independent of the accretion flow parameters. From the figure, one can clearly observe that the sonic points always occur at sub-Keplerian values of the angular momentum. This was shown earlier in the case of accretion around a non-rotating black hole using the PW80 potential (C96b). In Fig. 3b, we show the effect of viscosity on the specific angular momentum at the sonic points. The curves are drawn for different values of the viscosity parameter, $\alpha_\Pi=0, 0.15, 0.30, 0.45, 0.60$ (from top to bottom), for $a=0.5$ and fixed specific energy at the sonic point ($E_c=1.005$). The solid, dotted and long-dashed parts of the curves represent the saddle-type, nodal-type and spiral-type sonic points, respectively. Notice that, with the increase of $\alpha_\Pi$, the values of $l_c$ at a sonic point gradually decreases and more inner saddle-type sonic points are replaced by nodal-type sonic points which are in turn replaced by spiral-type sonic points. At high enough value of $\alpha_\Pi$, all the inner saddle-type sonic points are replaced by spiral-type sonic points and the accretion flow is destined to pass through the outer sonic point only. \par

In Fig. 4, we classify the parameter space available for stationary accretion solutions that allow multiple sonic points. In Fig. 4a, we show the parameter space spanned by the specific energy and the specific angular momentum of the accretion flow for spin parameters $a=0, 0.5, 0.9$ in the inviscid ($\alpha_\Pi=0$) limit. We find that the parameter space shifts towards lower angular momentum and higher energy. This is evident as the marginally stable angular momentum decreases for rapidly rotating black holes. In Fig. 4b, we show the parameter space for accretion flow that can pass through the inner sonic point, spanned by the specific energy at the inner sonic point and the specific angular momentum at the inner edge of the accretion disk, for viscosity parameters $\alpha_\Pi=0, 0.01, 0.05$ and spin parameter $a=0.5$. When viscosity is increased, the parameter space gradually shrinks and shifts towards lower angular momentum due to enhanced angular momentum transport. \par

\begin{figure}
    \plotone{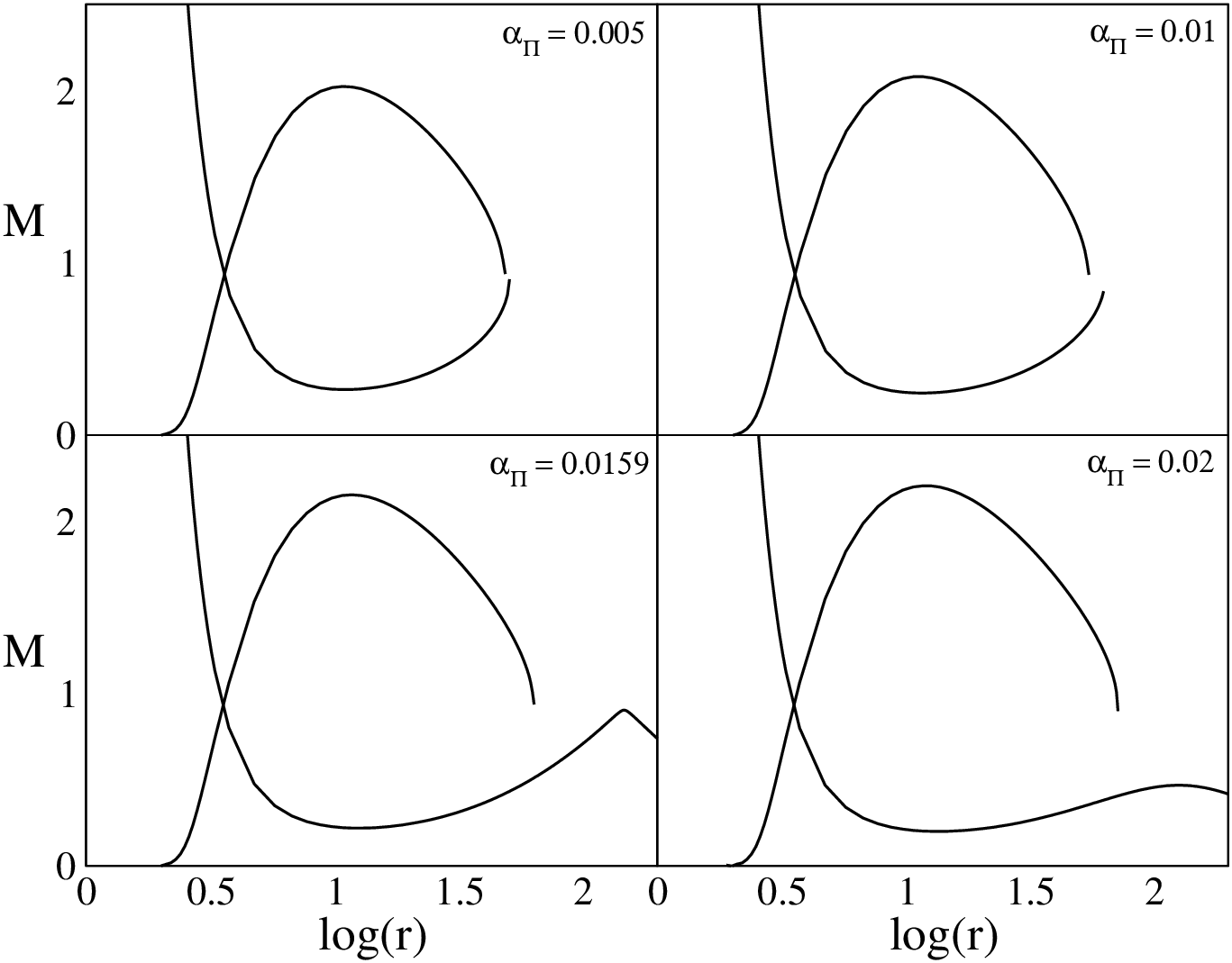}
    \caption{The Mach number variation with the radial coordinate for the parameters $a=0.5,\, l_{in}=3.0,\, E_{in}=1.003$ and different values of the viscosity parameter, $\alpha_\Pi=0.005, 0.01, 0.0159, 0.02$ for the flow to pass through the inner sonic point. Only half of the spirals are shown for clarity. With increase of $\alpha_\Pi$, the closed topology gradually opens up. For $\alpha_\Pi^c=0.0159$, the subsonic branch passing through the inner sonic point is clearly seen to pass through the outer sonic point also. For $\alpha_\Pi>\alpha_\Pi^c$, the flow passes only through the inner sonic point.}
\end{figure}

In Fig. 5, the effect of the viscosity on the solution topologies are shown for flows passing through the inner saddle-type sonic point. The parameters chosen are $a=0.5,\; l_{in}=3.0,\; E_{in}=1.003$ that correspond to a shock solution in accretion in the inviscid limit. The viscosity parameters are mentioned in each panel. We find that for small values of $\alpha_\Pi$, the topology is closed and the angular momentum of the flow can join a Keplerian disk only if a shock is formed. With increase in $\alpha_\Pi$, the flow topology gradually opens up. There exists, however, a critical value of the viscosity parameter, $\alpha_\Pi^c$, such that the subsonic branch passing through the inner saddle-type sonic point must also pass through the outer saddle-type sonic point for $\alpha_\Pi=\alpha_\Pi^c$ \citep{chakrabarti1990standing}. Thus, the value of $\alpha_\Pi^c$ depends on the flow parameters and it is the highest possible viscosity parameter for which the flow can pass through two saddle-type sonic points with the possibility of a shock transition in between them. In the present case, we find $\alpha_\Pi^c=0.0159$. When $\alpha_\Pi>\alpha_\Pi^c$, the flow leaves the accretion shock regime and joins with the Keplerian disk and enters the event horizon through the inner sonic point only. \par

\section{Shock Solutions} \label{sec:ss}
Black hole accretion is a transonic process. This means that the flow must pass through at least one saddle-type sonic point. However, in order to form a shock, the flow must possess two saddle-type sonic points. Initially, the accretion flow is subsonic and has a negligible radial velocity at a large distance from the black hole. the inward drift of matter allows the radial velocity to gradually increase and the flow becomes supersonic after passing through the outer sonic point. The flow then makes a discontinuous jump to the subsonic branch through a shock at a location where the Rankine-Hugoniot shock conditions are satisfied. The flow subsequently passes through the inner sonic point to become supersonic again before entering the event horizon of the black hole. \par

\begin{figure*}
    \epsscale{0.85}
    \plotone{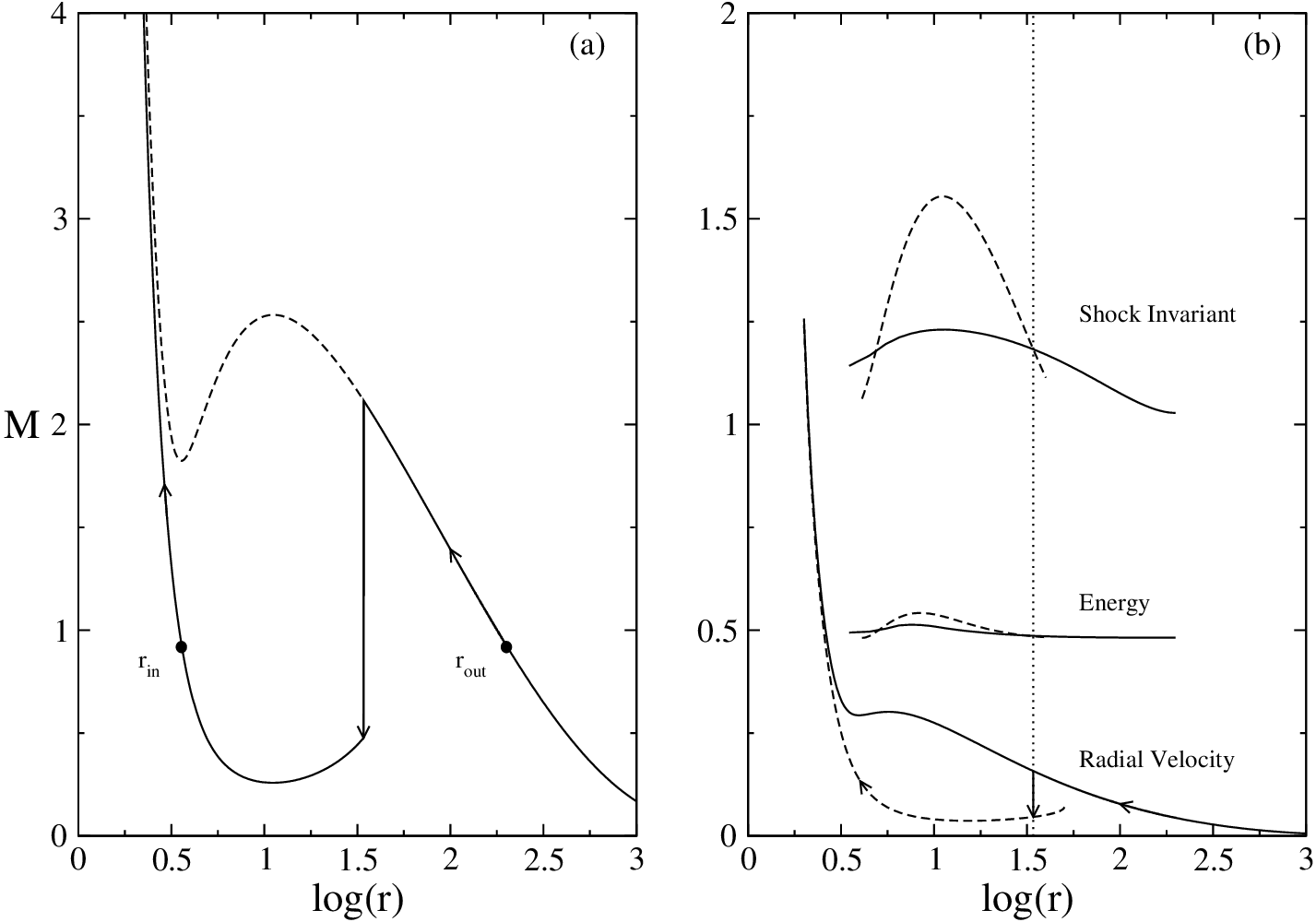}
    \caption{(a) The Mach number variation as a function of the radial coordinate for shock solution in accretion having the parameters $a=0.5, \alpha_\Pi=0.007, l_{in}=3.0$ and $r_{out}=200$. The Rankine-Hugoniot shock conditions are satisfied at $r_s=34.14$ (shown with a vertical arrow) and the subsonic post-shock flow subsequently becomes supersonic again at the inner sonic point located at $r_{in}=3.59$. (b) The variation of the shock invariant, the specific energy and the radial velocity of the accretion flow are shown as a function of the radial coordinate. The solid and dashed curves represent the supersonic branch passing through the outer sonic point and the subsonic branch passing through the inner sonic point, respectively. The location of the stable shock location is indicated by the dotted line.}
\end{figure*}

It is important to note that the existence of a shock is a consequence of the second law of thermodynamics. This is because a shock always connects two flow solutions, namely the supersonic branch of the pre-shock flow that passes through the outer sonic point and the subsonic branch of the post-shock flow that passes through the inner sonic point, having different entropies. In general, the entropy of the post-shock flow is higher than that of the pre-shock flow and this favours the formation of shocks in accretion flows. This situation is opposite in case of winds. Moreover, the shock front is assumed to be thin compared to the length scales in the pre-shock and post-shock flows, and the time for the matter to pass through the shock is short compared to the pre-shock and post-shock timescales. This allows to deduce the net impact of the shock on the flow without any reference to the detailed internal structure of the shock. We treat the shock as a discontinuous jump across which certain junction conditions, that enable us to relate the post-shock flow and its thermodynamic variables to their pre-shock counterparts, must be satisfied. \par

Unlike the case for an inviscid flow, where both the saddle-type sonic points can be obtained a priori, the analysis is more complex for viscous flows. This is because neither the specific energy, nor the specific angular momentum of the flow remains constant as in the case with inviscid flows. In what follows, we use the approach of C96b to find the shock locations. We use only three free parameters, namely the location of the outer sonic point $r_\mathrm{out}$, the specific energy $E_\mathrm{in}$ at the inner sonic point and the specific angular momentum $l_\mathrm{in}$ at the inner edge of the accretion disk for a certain value of $\alpha_\Pi$ to obtain a complete solution. We start by numerically integrating Equation (23) both outward and inward from the outer sonic point. This gives the flow topology that passes through the outer sonic point. However, this doesn't give the parameters of the subsonic branch that passes through the inner sonic point. So we choose the inner sonic point $r_\mathrm{in}$ arbitrarily. The flow passing through $r_\mathrm{out}$ will also pass through the inner sonic point $r_\mathrm{in}$ only if the Rankine-Hugoniot conditions, namely the conservation of local energy flux, mass flux and momentum flux, are satisfied at a location in between $r_\mathrm{in}$ and $r_\mathrm{out}$. We iterate $r_\mathrm{in}$ until we find the shock location $r_\mathrm{s}$ where the Rankine-Hugoniot shock conditions are uniquely satisfied. Across a shock, the entropy is discontinuous and the flow jumps from a supersonic low-entropy state to a subsonic high-entropy state. Moreover, the shock-invariant quantity
\begin{equation}
    C = \frac{[(3\gamma-1)M+(2/M)]^2}{2+(\gamma-1)M^2}
\end{equation}
is satisfied independently by the Mach numbers of the pre-shock and post-shock branch at the shock location. \par

\begin{figure}
    \plotone{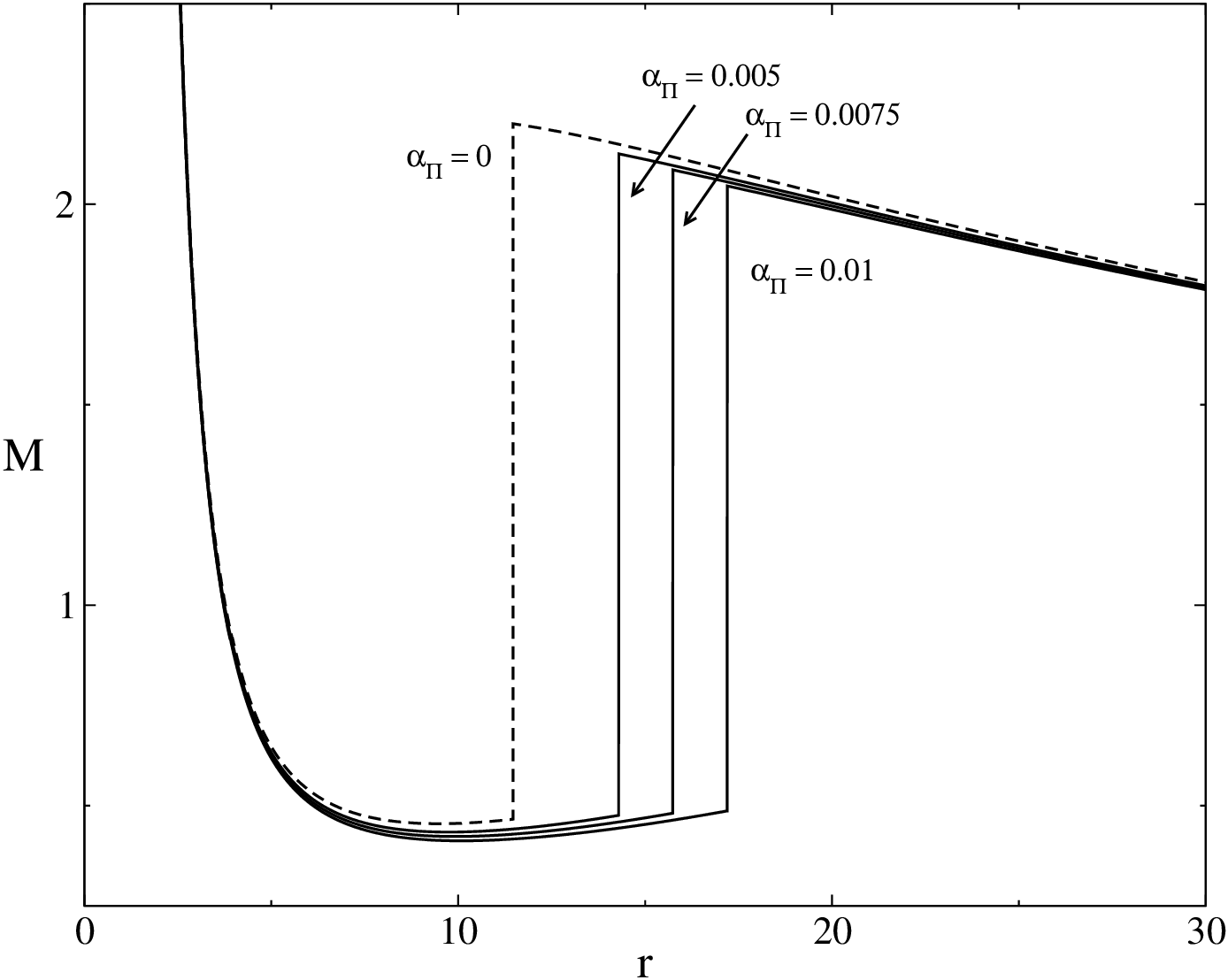}
    \caption{The effect of the viscosity on the location of the standing shock is illustrated. The parameters of the accretion flow are $a=0.5, l_{in}=2.9, E_{in}=1.0045$ and the viscosity parameters are $\alpha_\Pi=0, 0.005, 0.0075, 0.01$, as indicated. As the viscosity parameter is increased, the shock becomes weaker and shifts outwards.}
\end{figure}

In Fig. 6a, we present an example of a shock solution where the Mach number ($M=v/a_s$) of the accretion flow is plotted as a function of the radial coordinate. We choose the parameters $a=0.5,\; \alpha_\Pi=0.007,\; l_{in}=3.0$ and $r_{out}=200$. Then, we iterate the inner sonic point $r_{in}$ and find that the Rankine-Hugoniot shock conditions forces the accretion flow to have a shock transition at $r_s=34.14$ and passes through the inner sonic point located at $r_{in}=3.59$. Although there exist a shock-free solution (dashed curve) that passes through the outer sonic point, the flow chooses the subsonic branch (for $r<r_s$) after a shock transition at $r=r_s$ as a consequence of the second law of thermodynamics. In Fig. 6b, we show the shock invariant and specific energy variations of the subsonic branch (dashed curves) passing through the inner sonic point and the supersonic branch (solid curve) passing through the outer sonic point. Clearly, the curves intersect at two points that represents two possible shock locations. However, only the outer shock location is stable for accretion flows \citep{chakrabarti1993smoothed}. The jump in the radial velocity is also shown at the shock location. \par

We now discuss the effect of viscosity on the location of shocks in accretion flows. We begin by choosing a shock solution in the inviscid ($\alpha_\Pi=0$) limit having the parameters $a=0.5,\; l=2.9,\; E=1.0045$ which corresponds to a shock transition at $r_s=11.46$ having shock strength (the ratio of the pre-shock and post-shock Mach numbers) $\mathcal{S}=4.72$, the inner and outer sonic points located at $r_{in}=3.93$ and $r_{out}=119.46$, respectively. To examine the effect of viscosity, we now fix the outer sonic point at $r_{out}=119.46$ and iterate the inner sonic point for different values of the viscosity parameter. The conserved specific angular momentum $l=2.9$ in the inviscid case is now chosen to be $l_{in}=2.9$ for the viscous cases. In Fig. 7, we present the Mach number variations with the radial coordinate. The dashed curve represent the inviscid case. For the cases with viscosity, the chosen values of $\alpha_\Pi$ are (from left to right) $\alpha_\Pi=0.005, 0.007, 0.01$. We find that the shocks are weaker and shifts away from the black hole as the viscosity is enhanced, as expected. The calculated shock locations are $r_s=14.29, 15.74, 17.19$ having shock strengths $\mathcal{S}=4.46, 4.33, 4.20$ for $\alpha_\Pi=0.005, 0.007, 0.01$, respectively. The reason for this behavior is that the angular momentum transport rate in the post-shock flow is enhanced compared to the pre-shock flow as the viscosity is increased. The inner sonic point, however, drift towards the black hole. In the increasing order of viscosity, we find that the inner sonic points are located at $r_{in}=3.903, 3.888, 3.874$ for the viscous cases. When the viscosity is sufficiently high, we find that the Rankine-Hugoniot conditions are not satisfied and stable shocks do not form at all. Instead, the flow is destined to remain subsonic and Keplerian throughout the accretion disk and becomes supersonic only after passing through the inner sonic point close to the black hole. \par

\section{Discussions} \label{sec:d}
The existence of shocks in accretion flows around black holes is an essential ingredient of the TCAF paradigm of CT95. According to the TCAF model, the post-shock region (also known as the CENtrifugal pressure dominated BOundary Layer, or CENBOL) behaves as the Compton cloud which can oscillate if its cooling timescale roughly matches the infall timescale \citep{molteni1996resonance} and can cause the observed QPOs in black hole candidates \citep[e.g.,][and references therein]{chakrabarti2008evolution}. In this paper, we study the effect of viscosity on shocked accretion flows around Kerr black holes. We fix the outer sonic point and iterate the inner sonic point to find that the shock location moves farther out as the viscosity is increased. This behavior is consistent with numerical simulations performed for accretion around non-rotating black holes \citep{chakrabarti1995viscosity, lanzafame1998smoothed, lee2011quasi, giri2012hydrodynamic}. On the contrary, the shock would drift inward if the inner sonic point is kept fixed instead of the outer sonic point \citep{chakrabarti1990theory, mondal2014compton}. Recently, numerical studies of sub-Keplerian transonic accretion flows around black holes have been performed using a general relativistic numerical simulation code \citep{kim2017general, kim2019general}. These works demonstrated that shocks are captured, in one dimension, exactly where the theoretical shock locations were predicted by the vertical equilibrium model \citep{chakrabarti1996global}. They further show that, in two-dimensional simulations, presence of turbulence due to the centrifugal barrier pushes the shock outwards and may also change the topology of the flow solutions altogether. Magnetic fields are also believed to have significant effect on accretion flows around black holes. \citet{chakrabarti1990weber} studied all possible solution topologies of magnetized transonic flows around non-rotating black holes and examined the formation of MHD shock in such flows. In the recent numerical simulations of \citet{garain2020effects}, the influence of magnetic flux tubes on accretion flows around non-rotating black holes has been studied. They observed that in the presence of magnetic flux tubes, the magnetic pressure increases inside the CENBOL region that causes the shock to form at a larger radius. Additionally, the outflow is also significantly enhanced with increase in magnetic field strength. We expect these behaviors to persist even in Kerr geometry.

\section{Concluding Remarks} \label{sec:cr}
Within the context of astrophysical scenarios, black holes are expected to possess considerable angular momentum \citep{bardeen1970kerr}. This means that the spin of a black hole can significantly influence various astrophysical phenomena. Since the determination of the spin is intricately linked to our understanding of accretion disk physics, it is essential to incorporate the effects of spin in models of black hole accretion disks. \par 

In this work, we have studied viscous transonic accretion flows around Kerr black holes within the framework of the pseudo-Kerr \citep{bhattacharjee2022transonic} formalism. Specifically, our aim is to investigate the effects of viscosity on the various properties of transonic accretion flows. Depending on the initial flow parameters, accretion flows may allow multiple sonic points which is a necessary, but not sufficient, condition for the flow to form a shock wave. We find that viscosity significantly affects the topological properties of accretion flows at the sonic points. Consequently, the parameter space allowing multiple sonic points shrinks and shifts towards lower values on the $l_{in}$ scale, where $l_{in}$ is the angular momentum at the inner edge of the accretion disk. Moreover, a critical value ($\alpha_\Pi^c$) exists for the viscosity parameter which acts as an upper bound that may allow the formation of shock waves provided the Rankine-Hugoniot shock conditions are satisfied. These shocks are not merely any transient phenomena, rather they are essential ingredients in models of accretion flows that rely on a sub-Keplerian component in addition to the standard Keplerian component (as in CT95). We find that the shocks are weaker and forms farther away from the black hole as the viscosity is increased. According to the TCAF solution of CT95, the shock location directly gives the size of the post-shock region that behaves like a Compton cloud. Consequently, the resonance oscillation of the shocks changes the size of the Compton cloud significantly and this is associated with the QPOs observed in black hole candidates \citep{chakrabarti2000correlation, chakrabarti2015resonance}. This oscillation frequency is approximately inverse of the infall timescale $t_i\sim r_s^{3/2}$, where $r_s$ is the shock location, and its dependence on the spin parameter and viscosity is evident. Thus, the QPO frequencies are directly related to the shock locations and this may aid to fit data with the TCAF solution and extract the spin parameter of various black hole candidates. We have ignored the effects of cooling processes on accretion flows in this paper. Thus, it is of interest to explore how cooling affects the solution topologies and the formation of shocks in accretion flows in Kerr geometry. We will address these issues using our pseudo-Kerr formalism in the forthcoming papers.


\begin{acknowledgments}
The authors would like to thank the anonymous reviewer for valuable comments and suggestions. The authors acknowledge a grant of the ISRO sponsored RESPOND project (ISRO/RES/2/418/18-19). A.B. also acknowledges a grant towards a Junior Research Scientist position at ICSP from Govt. of West Bengal, India.\end{acknowledgments}

%





\bibliography{paper3}{}
\bibliographystyle{aasjournal}



\end{document}